\documentclass[letter]{aa} 

\usepackage{graphicx}
\usepackage{txfonts}
\usepackage{natbib}
\usepackage[colorlinks=true,allcolors=blue]{hyperref}
\newcommand{\orcid}[1]{\unskip\protect\href{https://orcid.org/#1}{\protect\includegraphics[width=8pt,clip]{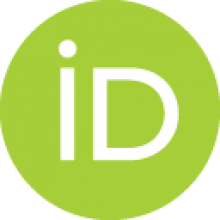}}}
\begin{document}

   \title{Discovery of a substellar companion in the TESS light curve of the $\delta$ Scuti/$\gamma$ Doradus hybrid pulsator HD\;31221}
    \titlerunning{Substellar companion of HD\;31221}
    \authorrunning{Sz. K\'alm\'an et al.}

   \author{Sz.~K\'alm\'an
          \inst{1,2,3,4}\orcid{0000-0003-3754-7889}
          \and
          A.~Derekas\inst{5,6,7}\orcid{0000-0002-6526-9444}
          \and 
          Sz.~Csizmadia\inst{8}\orcid{0000-0001-6803-9698}
          \and
          Gy.~M. Szab\'o\inst{2,5,6}\orcid{0000-0002-0606-7930}
          \and 
          V.~Heged\H{u}s \inst{3, 6}\orcid{0000-0001-7699-1902}
          \and
          A.~M.~S.~Smith \inst{8}\orcid{0000-0002-2386-4341}
          \and 
          J.~Kov\'acs \inst{2,5,6}\orcid{0000-0002-1883-9555}
          \and
          C.~Ziegler \inst{9}\orcid{0000-0002-0619-7639}
          \and
          A.~P\'al\inst{1,4}\orcid{0000-0001-5449-2467}
          \and 
          R.~Szab\'o\inst{1,4,10,11}\orcid{0000-0002-3258-1909}
          \and
          H.~Parviainen\inst{12,13}\orcid{0000-0001-5519-1391}
          \and
          F.~Murgas\inst{12,13}\orcid{0000-0001-9087-1245}
          }

\institute{ 
        Konkoly Observatory, Research Centre for Astronomy and Earth Sciences, ELKH, Konkoly-Thege Miklós út 15–17., H-1121, Hungary 
\and  
        MTA-ELTE Exoplanet Research Group, Szombathely, Szent Imre h. u. 112., H-9700, Hungary       
\and
        ELTE E{\"o}tv{\"o}s Lor\'and University, Doctoral School of Physics,  Budapest, Pázmány Péter sétány 1/A, H-1117, Hungary
\and 
        CSFK, MTA Centre of Excellence, Budapest, Konkoly Thege Miklós út 15-17., H-1121, Hungary
 \and
        ELTE E{\"o}tv{\"o}s Lor\'and University, Gothard Astrophysical Observatory, Szombathely, Szent Imre h. u. 112., H-9700, Hungary
\and    
        MTA-ELTE  Lend{\"u}let "Momentum" Milky Way Research Group, Hungary
\and 
    ELKH-SZTE Stellar Astrophysics Research Group, H-6500 Baja, Szegedi út, Kt. 766, Hungary  
 \and
        Deutsches Zentrum für Luft- und Raumfahrt, Institute of Planetary Research, Rutherfordtstrasse 2, D-12489 Berlin, Germany      
\and 
    Department of Physics, Engineering and Astronomy, Stephen F. Austin State University, 1936 North St, Nacogdoches, TX 75962, USA
\and
    MTA CSFK Lend\"ulet Near-Field Cosmology Research Group, Konkoly-Thege Miklós út 15–17., H-1121, Hungary
\and
    ELTE E\"otv\"os Lor\'and University, Institute of Physics, P\'azm\'any P\'eter s\'et\'any 1/A, H-1117 Budapest, Hungary
\and
    Dept. Astrof\'isica, Universidad de La Laguna (ULL), E-38206 La Laguna, Tenerife, Spain
\and
    Instituto de Astrof\'isica de Canarias (IAC), E-38200 La Laguna, Tenerife, Spain}

   \date{Received ...; accepted ...}

 
  \abstract
   {Close-in, sub-stellar companions to $\delta$ Scuti type stars present a highly suitable testbed for examining how planetary-mass objects can influence stellar pulsations.}
   {We aim to constrain the mass of HD 31221 b, probe its atmosphere, and demonstrate how it affects the pulsational pattern of its host, HD 31221.}
   {We made use of the available data from the short-cadence Transiting Exoplanet Survey Satellite (TESS). We modeled the nine observed transits and the out-of-phase variations, including Doppler beaming, ellipsoidal variations, and the reflection effect. We also incorporated ground-based photometry from the MuSCAT2 imager installed at the 1.52 m Telescopio Carlos Sanchez in the Teide Observatory, Spain, as well as speckle interferometry from the Southern Astrophysical Research telescope.}
   {We found HD 31221 b to have an orbital period of $4.66631 \pm 0.00011$~days, with a radius of $1.32 \pm 0.14$~R$_J$ and a mass of $11.5 \pm 10.3$~M$_J$ (from the ellipsoidal effect),  making it consistent with either a brown dwarf or a giant planet. As HD 31221 is a rapid rotator ($v \sin I_\star  = 175.31 \pm 1.74$~km~s$^{-1}$), we deduced the spin-orbit misalignment to be $\lambda = -121.6 \pm 14.4^\circ$ and $I_\star = 55.9 \pm 11.3^\circ$. The phase curve is dominated by the reflection effect, with a geometric albedo of $1.58 \pm 0.50$. We also found evidence that HD 31221 is a $\delta$ Scuti/$\gamma$ Doradus hybrid pulsator. There are three cases for which the $3$rd, $85$th, and $221$st orbital harmonics almost exactly coincide with peaks in the Fourier spectrum of the star, hinting at tidally perturbed stellar oscillations.}
    {HD 31221 b is the third substellar object that is found to be disrupting the pulsations of its host, following HAT-P-2 and WASP-33. Additional photometric observations by CHEOPS and/or PLATO can be used to further constrain its mass and provide a more in-depth analysis of its atmosphere.}


   \maketitle
%

\section{Introduction}

There are only a handful of known exoplanets orbiting $\delta$ Scuti type stars \citep{2021AJ....162..204H}, with the most famous examples being hot Jupiters KOI-976 b \citep{2019AJ....158...88A} and WASP-33 b \citep{2006MNRAS.372.1117C, 2010MNRAS.407..507C, 2011A&A...526L..10H, 2020A&A...639A..34V}. There is also a known brown dwarf orbiting the $\delta$ Scuti star Chang 134, discovered via pulsation timing  \citep{2018haex.bookE...6H, 2022arXiv220901220V}. Planets and brown dwarfs with short orbital periods around A-F stars generally serve as excellent testbeds for atmospheric analyses through the study of phase curves, as in the case of KELT-1 b \citep{2012ApJ...761..123S,2020AJ....160..211B, 2021A&A...648A..71V, 2021AJ....162..127W,  2022arXiv220903890P} and  KELT-9 b \citep{2017Natur.546..514G, 2022A&A...666A.118J}. Such studies are enabled by the space-based photometry with \textit{Spitzer}/IRAC,  Transiting Exoplanet Survey Satellite \citep[TESS;][]{2015JATIS...1a4003R}, and CHaracterising ExOPlanet Satellite \citep[CHEOPS;][]{2021ExA....51..109B}.

In this letter, we present the discovery of a substellar companion of HD 31221 (HIP 22838, TIC 68573534, TOI-4597, Gaia DR2 3412431441720096128; $\alpha_{\rm J2000} = 4^{\rm h}54^{\rm m}49.39^{\rm s}$, $\delta_{\rm J2000} = +22^\circ 08' 47.35''$). In Sectors
43 and 44, TESS observed the bright A2 star ($V= 8.02$ mag) that is rapidly rotating ($v \sin I_\star = 175.31 \pm 1.74$~km~s$^{-1}$, as derived from observations by the Tillinghast Reflector Echelle Spectrograph, TRES, mounted on the 1.5-Meter telescope of the Fred Lawrence Whipple Observatory, FLWO). The light curve of HD 31221 is dominated by stellar oscillations, which we attribute to both $\gamma$ Doradus and $\delta$ Scuti type pulsations. There are also nine transits observed in the two adjacent Sectors, corresponding to an orbital period of $\approx 4.7$~days. A mass estimation for HD 31221 b from radial velocity observations is not feasible because of the rapid rotation and pulsation. We therefore rely on the modeling out-of-transit variations, including Doppler beaming and ellipsoidal variations \citep{2007ApJ...670.1326Z, 2011MNRAS.415.3921F}, to constrain its mass and the reflection effect to constrain its geometric albedo.

This letter is structured as follows. In Sect. \ref{sec:meth}, we describe the light curve preparation and the models used in our analysis. In Sect. \ref{sec:params}, we present the parameters that can be determined through the light curve analysis. In Sect. \ref{sec:osc}, we briefly analyze the possibility that HD 31221 b is influencing the pulsations of its host through tidal interactions.

\section{Methods} \label{sec:meth}

\begin{figure}
    \centering
    \includegraphics[width = \columnwidth]{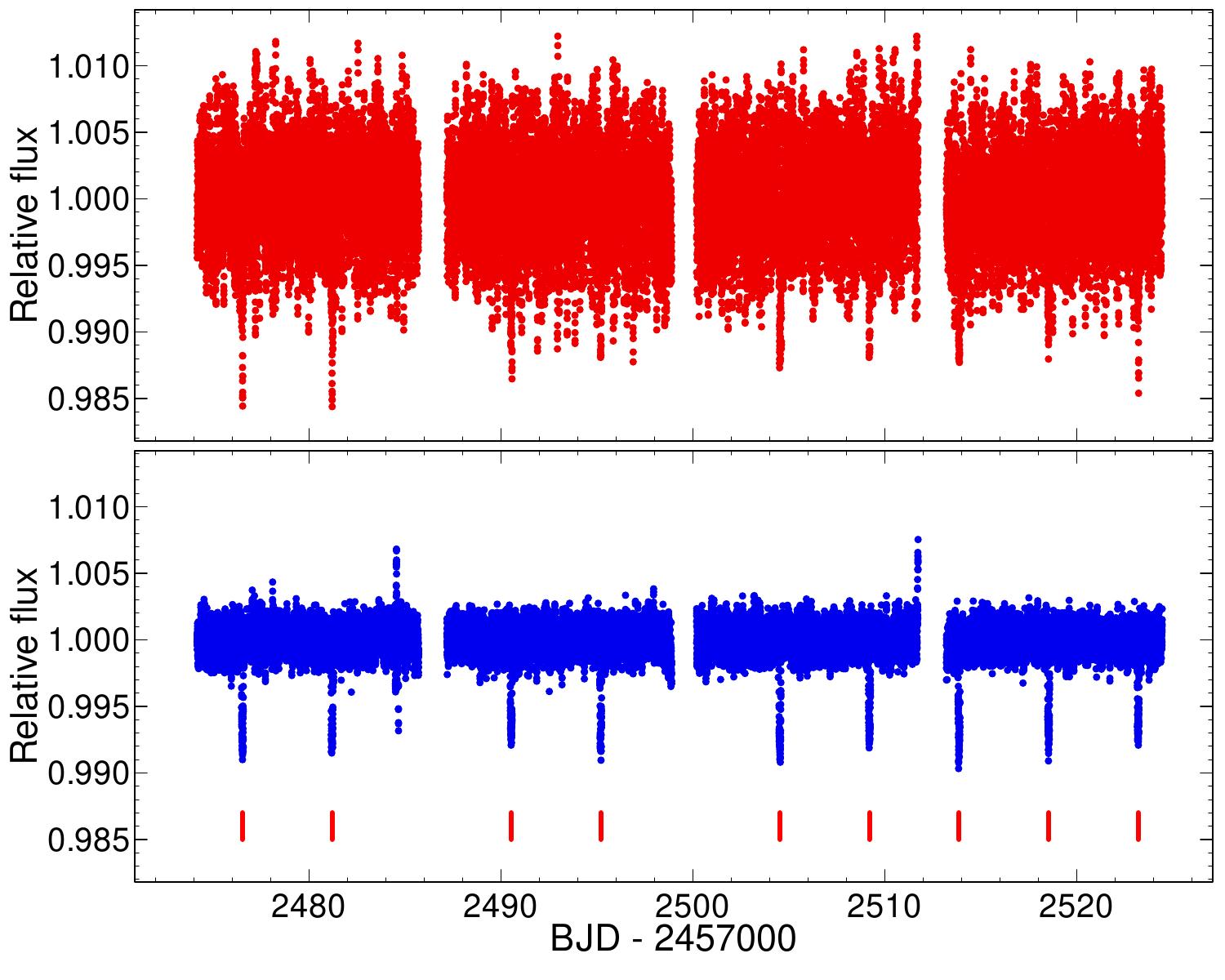}
    \caption{SAP lightcurve of HD 31221 (top panel). Removing the signal produced by the first $80$ frequencies with the highest amplitudes from the raw LC yields the input light curve (bottom panel) to our fitting. The nine observed transits are marked by red ticks in the bottom panel.}
    \label{fig:inputlc}
\end{figure}
\subsection{Light curve preparation}
TESS observed HD 31221 in Sectors 43 and 44 (GI proposal: G04106 -- Huber, D). We obtained the two-minute exposure-time Simple Aperture Photometry (SAP) light curves (LCs), using the \texttt{lightkurve} software package \citep{lightkurve}, while making use of the \texttt{astropy} package \citep{astropy:2013, astropy:2018, astropy:2022} and \texttt{astroquery} \citep{2019AJ....157...98G}. We removed all data points with a non-zero quality flag. The combined LCs from the two sectors are plotted in the upper panel of Fig. \ref{fig:inputlc}. The light curve shows $\delta$ Scuti type oscillations.

In order to fit the transit light curve, we subtracted the pulsation signal from the light curve. We used \texttt{period04} \citep{2005CoAst.146...53L} to identify the frequencies of the oscillations. First, we masked the transits, then computed the Fourier spectrum and removed the $80$ highest amplitude components from the observed light curve. This pre-cleaning procedure yielded the light curve shown in the bottom panel of Fig. \ref{fig:inputlc} and provided a sufficiently high S/N ratio for the analysis in Sect. \ref{sec:lcmod}. The result of the more detailed frequency analysis is discussed in \ref{sec:osc}.

\subsection{Origin of the transit}

\begin{figure}
    \centering
    \includegraphics[width = \columnwidth]{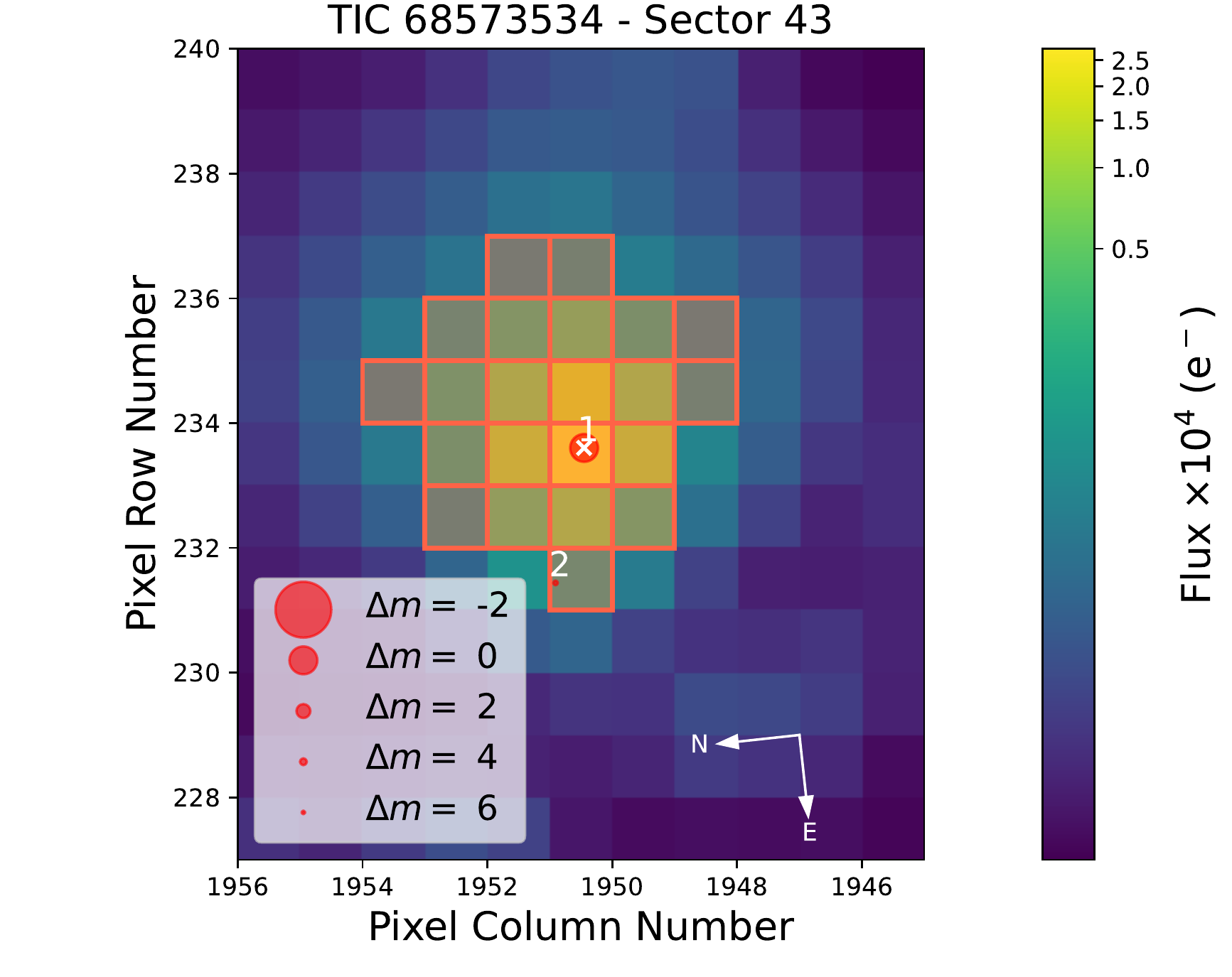}
        \caption{Results of \texttt{tpfplotter} showing HD 31221, the aperture used for the photometry and the only star that is known to be located in the aperture within $\Delta m = 6$ magnitudes. The TESS pixel scale is $21$''.}
    \label{fig:tpf}
\end{figure}

As HD 31221 is both rapidly rotating ($v \sin I_\star =  175.31 \pm 1.74$~km~s$^{-1}$) and exhibiting stellar oscillations, using radial velocity measurements to derive the mass of the object causing the transit signals (Fig. \ref{fig:inputlc}, lower panel) is not feasible. To exclude the possibility that a background eclipsing binary is causing these features, we made use of \texttt{tpfplotter} \citep{2020A&A...635A.128A} to detect all stars that are near the target (based on the Gaia DR3 catalog). Figure \ref{fig:tpf} suggests that there is only one known star in the default aperture used in the TESS pipeline within six magnitudes of HD 31221. The contaminating star, Gaia DR3 3412431471783190016 ($\alpha_{\rm J2000} = 4^{\rm h}54^{\rm m}52.22^{\rm s}$, $\delta_{\rm J2000} = +22^\circ 09' 1.23''$), marked by the number $2$ in Fig. \ref{fig:tpf}, is $G = 5.08$ magnitudes fainter than HD 31221. To exclude the possibility that the transit signal (Fig. \ref{fig:inputlc}) originates from Gaia DR3 3412431471783190016, we created two custom apertures to separate the point spread function (PSF) of the two stars for measurements taken during Sector 43 (Fig. \ref{fig:aps}): one where (most of) the signal coming from the contaminating star is excluded and another where (most of) the flux originating from HD 31221 is excluded. The resultant light curves (Fig. \ref{fig:conta}) suggest the transit signals are unrelated to Gaia DR3 3412431471783190016. 

As an additional test, we observed a light curve of Gaia DR3 3412431471783190016 covering the ingress of the transit with the MuSCAT2 multicolor imager \citep{Narita2018} installed at the 1.52~m Telescopio Carlos Sanchez (TCS) in the Teide Observatory, Spain. The observations were carried out simultaneously in the $g$, $r$, $i$, and $z_\mathrm{s}$ bands in good weather conditions on the night of February  18, 2023 from 20:30 to 23:30 UT, and the photometry was carried out with the MuSCAT2 photometry pipeline described in \citet{Parviainen2019}. The observations securely reject Gaia DR3 3412431471783190016 as the possible source of the signal, as can be seen from Fig.~\ref{fig:muscat-gaia}. As the stellar pulsations of HD 31221 have a comparable amplitude to the transit depth of its companion and the fact that only the ingress of HD 31221 b would have been observable on the night of 2023 Feb 18, the MuSCAT2 observations of the proposed host yielded only a tentative transit detection.

We therefore suggest that the probability of the transit signals having an origin outside of HD 31221 is extremely low. We also searched for companions with speckle interferometry by the Southern Astrophysical Research (SOAR) telescope \citep{2018AJ....155..235T} in the Cousins $I$-band on November 4, 2022\footnote{https://exofop.ipac.caltech.edu/tess/view\_tag.php?tag=422227}. The sensitivity curve (Fig. \ref{fig:soar}) suggests that there are no stars within $3$'' and $\Delta m = 7$ magnitudes, excluding the possibility of a blended star and, ultimately, further reducing the probability that the observed signal originates anywhere other than the HD 31221 system.

\subsection{Light curve analysis} \label{sec:lcmod}

We used the Transit and Light Curve Modeler \citep[\texttt{TLCM};][]{2020MNRAS.496.4442C} to analyze the light curve. In \texttt{TLCM}, the transits are fitted using a Mandel-Agol model \citep{2002ApJ...580L.171M}, described by the orbital period, $P$, the time of midtransit, ($t_C$), the impact parameter, ($b = a/R_S \cos i$, with $i_p$ being the orbital inclination relative to the of sight), the planet-to-star radius ratio ($R_P/R_S$), and the scaled semi-major axis ($a/R_S$). We made use of a quadratic limb-darkening law characterized by the coefficients $u_+ = u_{\rm linear} + u_{\rm quadratic}$ and $u_- = u_{\rm linear} - u_{\rm quadratic}$, which were left as free parameters of the fit. We assumed a circular orbit. Based on the TESS Input Catalog\footnote{https://tess.mit.edu/science/tess-input-catalogue/}, we used $T_{\rm eff} = 7712 \pm 240$ K, $\log g = 4.31 \pm 0.08$ and $R_S = 1.57 \pm 0.05$ (with the latter used as a prior). The stellar parameters were taken into account via the empirical formulae of \cite{2011MNRAS.417.2166S}, with the assumption of solar-like metallicity.

Rapid rotation of the host star causes asymmetric transits as the transit chord of HD 31221 b crosses in front of the cooler equator and hotter poles of its host  \cite[see e.g.,][]{2009ApJ...705..683B}. This effect is known as gravity darkening. \texttt{TLCM} has a built-in gravity-darkening model \citep[see][for details]{2020A&A...643A..94L, 2020MNRAS.496.4442C}, which allowed us to constrain the inclination of the stellar rotational axis, $I_\star$, and the projected spin-orbit misalignment, $\lambda$. These parameters depend on the gravity darkening exponent $\beta$, which describes the surface brightness distribution. During our analysis, we fixed $\beta = 0.25$, based on the theoretical calculations of \cite{1924MNRAS..84..665V} and fit $I_\star$ and $\Omega_\star = \lambda + 90^\circ$.

We also modeled the out-of-transit variations, including Doppler beaming, the ellipsoidal effect, and the reflection effect \citep{2007ApJ...670.1326Z, 2011MNRAS.415.3921F}. This approach is an established way to confirm the nature of exoplanets \citep[see e.g. Kepler-41 b,][]{2013ApJ...767..137Q}. Through the Doppler beaming (characterized by the semi-amplitude of the radial velocity curve, $K$) and the ellipsoidal effect (described by the photometric mass, $q_{\rm ell}$), we were able to constrain the mass of the planets and brown dwarfs orbiting the host stars. The coefficients of these two effects are calculated based on the stellar parameters and they were not fitted independently of each other \citep{2020MNRAS.496.4442C}. The phase-curve variations were taken into account via a Lambertian phase function:
\begin{equation}
    \frac{F_{\rm ph}}{F_S} = \frac{I_P}{I_S} \left( \frac{R_P}{R_S} \right)^2 + A_g  \left( \frac{R_P}{R_S} \frac{R_S}{d} \right)^2 \frac{\sin \alpha - \alpha \cos \alpha}{\pi},
\end{equation}
where $F_{\rm ph}$ and $F_{S}$ are the reflected and stellar fluxes, respectively, $I$ represents the passband-specific intensity, $A_g$ is the geometric albedo, and $d$ is the mutual star-planet distance. The phase angle $\alpha$ can be described by
\begin{equation}
    \cos{\left(\alpha +\varepsilon \right)} = \cos{\left(\omega + \nu\right)} \sin{i_p},
\end{equation}
where $\varepsilon$ is the shift of the brightest point of the planet from the substellar point, and $\nu$ and $\omega$ are the true anomaly and the argument of the periastron, respectively \citep{2021arXiv210811822C}.

In order to handle the remaining signals caused by stellar pulsations and any instrumental effects, we used the wavelet-based routines of \citep{2009ApJ...704...51C} built into\texttt{TLCM}. These allowed us to fit the correlated noise (characterized by $\sigma_r$ and $\sigma_w$ for the red and white components, respectively) simultaneously with the signals caused by HD 31221 b. This approach for handling the correlated noise was tested on synthetic light curves by \citep{2021arXiv210811822C} and \cite{2022arXiv220801716K}, and it was found to be consistent. Fitting the combination of gravity darkening and stellar oscillations in this way was also tested on WASP-33 \citep{2022A&A...660L...2K}, where it was found to yield parameters that are compatible with other photometric studies \citep{2020A&A...639A..34V,2022ApJ...925..185D} and Doppler-tomography \citep[e.g.][]{2021A&A...653A.104B}. We also included a height correction \citep[c.f. Eq (47) of][]{2020MNRAS.496.4442C}, denoted by $h$.

\section{Results}

\subsection{Parameters of HD 31221 b} \label{sec:params}
\begin{figure}
    \centering
    \includegraphics[width = \columnwidth]{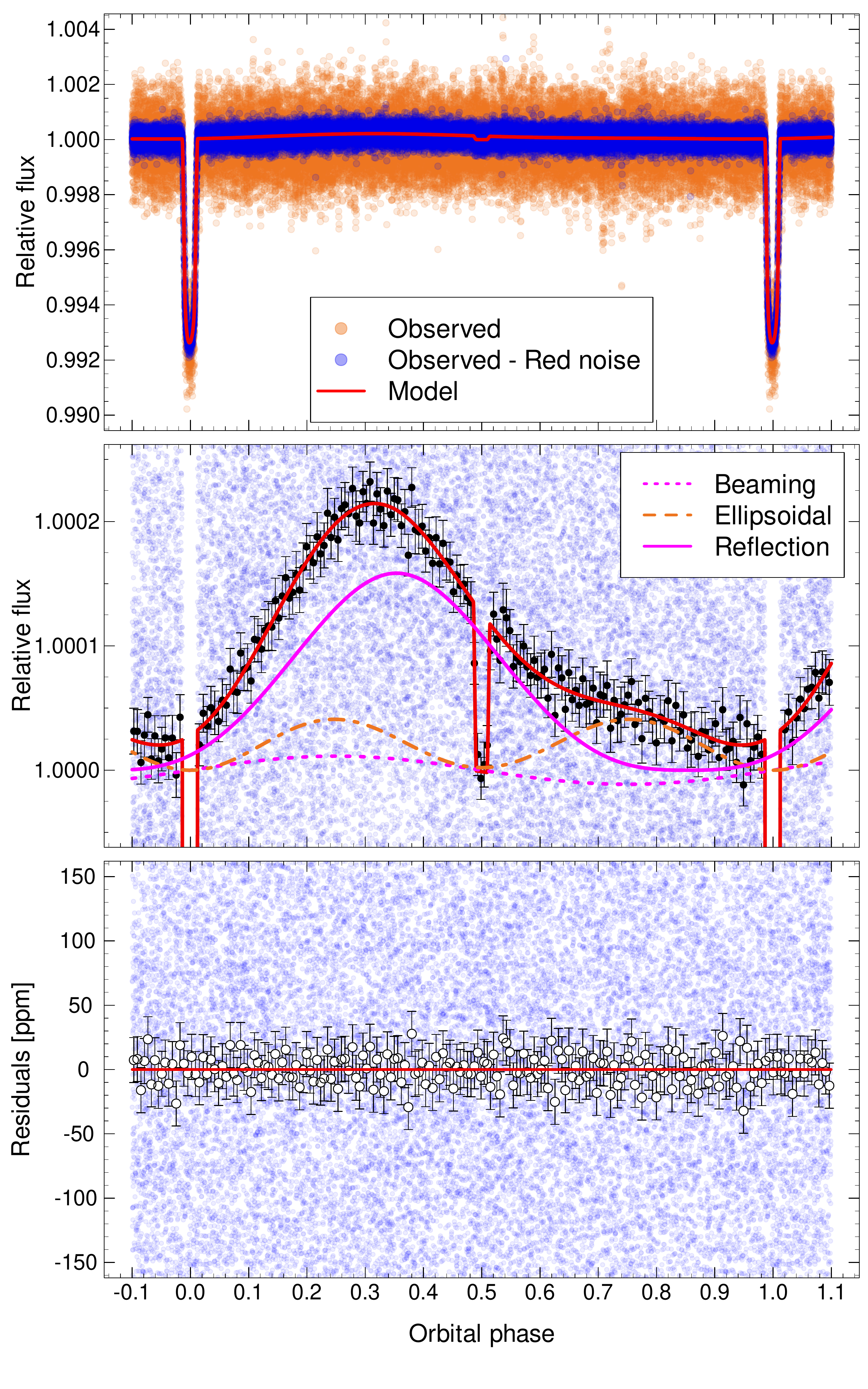}
    \caption{Phase-folded, pre-cleaned light curve of HD 31221 b (top panel, orange). The noise-corrected light curve is shown with blue dots on the top and middle panels. The solid red line represent the best-fit model. Black dots and the corresponding error bars of the middle panel represent $200$-points bins. The out-of-transit variations are dominated by the reflection effect (middle panel, solid magenta line), but the ellipsoidal effect (dashed orange line, middle panel) and the Doppler-beaming (dashed magenta line, middle panel) are also detectable.The residuals are plotted on the bottom panel with a similar scale to the middle panel. White circles with black outlines represent the binned data residuals with the same binning as on the middle panel. The respective error bars are shown with black. The orange dots of the top panels represent the same LC as on the lower panel of Fig. \ref{fig:inputlc}.}
    \label{fig:phase}
\end{figure}

Solving the light curves from the two adjacent sectors yielded the results seen in Table \ref{tab:params}. The phase-folded best-fit solution is shown in Fig. \ref{fig:phase}. Based on the $T_{\rm eff}$ -- $R_S$ empirical calibration \citep{2011MNRAS.417.2166S}, we estimated the following stellar parameters using \texttt{TLCM}: $R_S = 1.550 \pm 0.060$~R$_\odot$ and $M_S = 1.447 \pm 0.028$~M$_\odot$ (see \cite{2020MNRAS.496.4442C} for more details about the absolute parameter estimations). Using our fitted $R_P/R_S$ ratio, we derive an absolute radius for the companion of $R_P = 1.32 \pm 0.14$~R$_J$. We can constrain the absolute mass of the object in two ways. From $q_{\rm ell} = \frac{M_P}{M_S}$, we get $M_P = 11.5 \pm 10.3$~M$_J$. By modeling the Doppler beaming, we can fit the radial velocity semi-amplitude ($K$). We can then express the true mass ratio $q$ via
\begin{equation}
    q = \frac{K}{\frac{2 \pi}{P}a\sin i_p -K}.
\end{equation}
Therefore, with its orbital period of $4.66631$ days, and a semi-major axis of $0.0618 \pm 0.0018$ AU, HD 31221 b has a mass of $M_P = 13.0 \pm 13.3$~M$_J$. Estimating the mass of HD 31221 b from these approaches yields values that are in good agreement. These $M_P$ values mean that HD 31221 b may  either be a planet or a brown dwarf. Based on these mass calculations, we can place a $1-\sigma$ upper limit of $26.3$~M$_J$ on its mass. 

We measured a high geometric albedo of $1.58 \pm 0.50$, corresponding to the reflection-dominated out-of-phase variations seen in the middle panel of Fig. \ref{fig:phase}. This results is in line with KELT-1 b's high dayside albedo \citep{2020AJ....160..211B}. The best-fit nightside brightness ratio of $0.0026 \pm 0.0035$ is consistent with $0$. We  derive an offset of the brightest point on the companions surface from the sub-stellar points to be $52.6 \pm 41.2^\circ$, suggesting the presence of hazes in the atmosphere. We also measured a significant secondary occultation depth ($99 \pm 9$ ppm, Fig. \ref{fig:phase}) applying a so-called Welch statistics \citep{10.2307/2332579}. The results of this test suggest that the occultation depth is statistically significant at $S = 12.9$  at a level of $p=4 \cdot 10^{-36}$.



It is also known that the beaming and reflection effects are degenerate  \citep{2020MNRAS.496.4442C}. Given the lack of reliable RV data, it is difficult to break this degeneracy. Also given the rapid rotation of the star and its oscillations, obtaining viable RV measurements is highly improbable. Nonetheless, in order to try to break this degeneracy, we also solved the LC by applying a Gaussian prior of $\mathcal{N}(0.5, 0.1)$ to $A_g$. The resulting parameters are also listed in Table \ref{tab:params}. Upon introducing this constraint, we get a lower value for the geometric albedo of $0.53 \pm 0.12$, but all other parameters are consistent within the uncertainties with leaving $A_g$ as a free parameter. For the mass of HD 31221 b in the case when the prior is applied on the geometric albedo, we get $M_P = 13.0 \pm 11.1$~M$_J$ (from the ellipsoidal variations) and $M_P = 15.6 \pm 14.2$~M$_J$ (from the Doppler beaming). Both of these values are in good agreement with each other and the mass estimates of the case when $A_g$ is left as a free parameter as well. When we apply a prior on $A_g$, the resultant masses are marginally higher. The derived parameters for both cases (with free $A_g$ and with a Gaussian prior on $A_g$) are also listed in Table \ref{tab:params}. We also checked for degeneracies between the fitted parameters by plotting the posteriors of the Markov chain Monte Carlo (MCMC) analysis using the routines of \cite{corner} for the case where $A_g$ was treated as a free parameter of the fit. The resultant corner plot is shown on Fig. \ref{fig:corner}. There is a clear degeneracy between the parameters used for the mass estimation, $K$ and $q_{\rm ell}$, expressed by the Pearson's $r$ value of $0.88$. This is not unexpected, as the two parameters are fitted jointly. There is also an apparent contradiction between the significant offset of the peak of the reflection effect (Fig. \ref{fig:phase}) and the large uncertainty with which the $\varepsilon$ parameter is described ($52.6^\circ \pm 41.2^\circ$ and $48.5^\circ \pm 57.6^\circ$ in the two tested cases). There are no significant correlations between $\varepsilon$ and any other parameters, with $r = -0.47$ being the highest value when comparing its distribution to $t_C$. We suggest two possible explanations for this anomaly, each of which can be tested by upcoming observations: (i)  the remaining quasi-periodic stellar oscillations influence the determination of $\varepsilon$, or (ii) HD 31221b experiences weather whereby the offset changes on timescales shorter than the two TESS sectors. The former assumption of these is unexpected \citep[c.f.][]{2020MNRAS.496.4442C}, while the latter can not be tested on the TESS data alone, as the slicing of the original light curve yields data where there are too few points to have a good enough signal-to-noise ratio (S/N) to reliably estimate $\varepsilon$.

\begin{table}
\caption{Best-fit parameters of HD 31221 b with the two considered cases: $A_g$ left as a completely free parameter and with a $\mathcal{N}(0.5, 0.1)$ Gaussian prior applied to it. The uncertainties correspond to $1\sigma$.}
\label{tab:params}
\centering
\scriptsize
\begin{tabular}{l l c c}
\hline
\hline
Parameter & Searchbox & Value & Value\\
 & & \textit{Free} $A_g$ & $A_g$ \textit{with prior} \\
\hline
$a / R_S$ &  [$5.0$, $10.0$] & $8.59 \pm 0.25$ & $8.58 \pm 0.24$ \\ 
 $R_P / R_S$ & [$0.0$, $0.2$] & $0.0863 \pm 0.0027$ & $0.0862 \pm 0.0027$ \\ 
 $b$ & [$0.0$, $1.0$] & $0.770 \pm 0.019$ & $0.770 \pm 0.018$ \\ 
 $P$ [days] &  [$4.63$, $4.69$]& $4.66631 \pm 0.00011$ & $4.66632 \pm 0.00010$ \\ 
 $t_C$ [BTJD] & [$2476.52$, $2476.54$] & $2476.5375 \pm 0.0014$ & $2476.5373 \pm 0.0014$\\ 
$u_{+}$ & [$-1.0$, $2.0$] & $0.80 \pm 0.16$ & $0.80 \pm 0.16$ \\ 
$u_{-}$ & [$-1.5$, $1.5$] & $-0.44 \pm 1.02$ & $-0.49 \pm 0.97$\\ 
 $\sigma_{r}$ [$100$ ppm] & [$0.0$, $10000.0$] & $665.00 \pm 2.94$ & $664.90 \pm 2.76$ \\ 
 $\sigma_{w}$ [$100$ ppm] & [$0.0$, $6000.0$] & $3.036 \pm 0.019$  & $3.036 \pm 0.018$ \\ 
$I_\star$  [$^\circ$]& [$0.0$, $180.0$] & $55.9 \pm 11.3$  & $55.9 \pm 11.1$ \\ 
$\Omega_\star$  [$^\circ$]& [$-180.0$, $180.0$] & $-31.6 \pm 14.4$  & $-30.9 \pm 14.6$ \\ 
$A_g$ & [$0.0$, $2.0$] & $1.58 \pm 0.50$ & $0.53 \pm 0.12$\\ 
$f$ & [$0.0$, $1.0$] & $0.0026 \pm 0.0035$ & $0.0035 \pm 0.0044$\\ 
$\varepsilon$  [$^\circ$]& [$-180.0$, $180.0$] & $52.6 \pm 41.2$ & $48.5 \pm 57.6$\\ 
$K$  [m s$^{-1}$]& [$0.0$, $6000.0$] & $1233 \pm 1242$ & $1463 \pm 1320$ \\ 
$q_{\rm ell}$ & [$0.0$, $0.2$] & $0.0076 \pm 0.0068$ & $0.0086 \pm 0.0073$\\  $h$ & [$-0.5$, $0.5$] & $-0.00031 \pm 0.00018$ & $-0.00028 \pm 0.00017$ \\
\hline
\multicolumn{4}{c}{Derived parameters} \\
\hline
\multicolumn{2}{l}{$M_S$ [M$_\odot$]} & $1.447 \pm 0.028$ & $1.449 \pm 0.028$\\
\multicolumn{2}{l}{$R_S$ [R$_\odot$]} & $1.550 \pm 0.060$ & $1.553 \pm 0.059$\\
\multicolumn{2}{l}{$R_P$ [R$_J$]} & $1.32 \pm 0.14$ & $1.32 \pm 0.12$\\
\multicolumn{2}{l}{Ellipsoidal mass [M$_J$]} & $11.5 \pm 10.3$ & $13.0 \pm 11.1$\\
\multicolumn{2}{l}{Beaming mass [M$_J$]} & $13.0_{-13.0}^{+13.3}$ & $15.6 \pm 14.2$\\
\multicolumn{2}{l}{$i$ [$^\circ$]} & $84.86 \pm 0.20$ &  $84.85 \pm 0.19$\\
\hline
\end{tabular}
\end{table}

Gravity darkening of HD 31221 causes asymmetric transits (Fig. \ref{fig:gravdark}). By modeling this effect, we were able to derive values for the inclination of the stellar rotational axis and the projected spin-orbit misalignment. There are four equivalent ($I_\star$, $\lambda$) pairs that cannot be distinguished from each other via light curve analyses: ($55.9 \pm 11.3 ^\circ$, $-121.6 \pm 14.4^\circ$), ($55.9 \pm 11.3 ^\circ$, $301.6\pm 14.4^\circ$), ($124.1 \pm 11.3 ^\circ$, $121.6 \pm 14.4^\circ$), and ($124.1 \pm 11.3 ^\circ$, $58.4 \pm 14.4^\circ$). To better constrain these angles, either photometric observations with a higher precision, or (in the case of $\lambda$) Doppler tomographic investigations are necessary. Using these angles, we can also derive the true spin-orbit angle \citep{2009ApJ...696.1230F}:
\begin{equation}
\varphi = \arccos \left( \cos I_\star \cos i_p + \sin I_\star \sin i_p \cos \lambda \right).  
\end{equation}
We find $\varphi = 112.5 \pm 11.9^\circ$, $61.2 \pm 11.9^\circ$, $118.8 \pm 11.9^\circ$, and $67.5 \pm 11.8^\circ$ in the four listed scenarios, respectively. Two of these suggest a near-polar orbit, as in the cases of, for example,  KELT-9 b \citep{2020AJ....160....4A}, MASCARA-1 b \citep{2017A&A...606A..73T, 2022A&A...658A..75H}, WASP-33 b \citep{2022ApJ...925..185D,2022A&A...660L...2K}.

\subsection{Stellar oscillations} \label{sec:osc}

\begin{figure*}
    \centering
    \includegraphics[width = \textwidth]{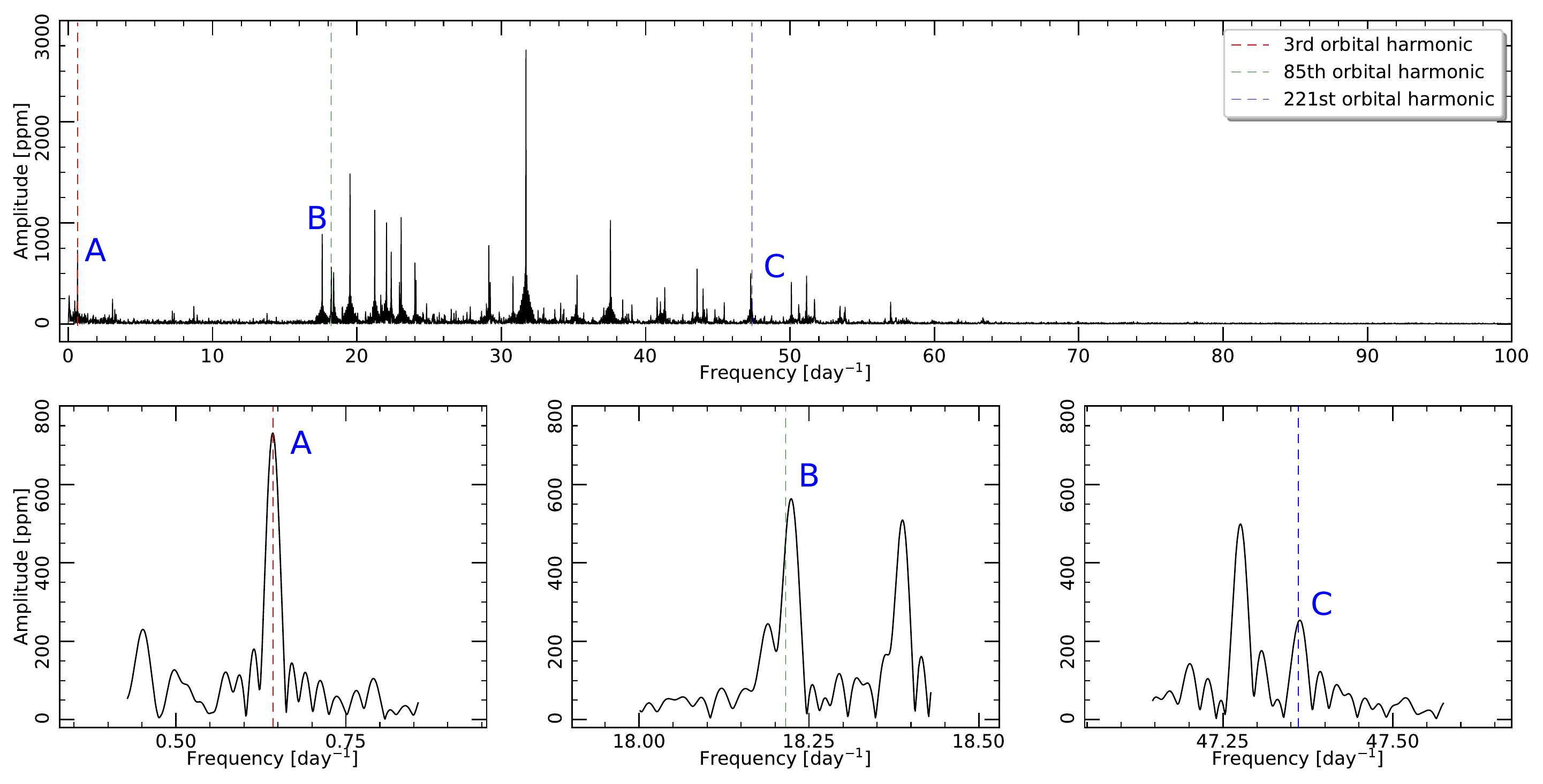}
    \caption{Fourier spectrum of HD 31221 (top panel). Portions of the spectrum around the three resonance-like features denoted by A, B, and C correspond to the $3$rd, $85$th and $221$st orbital harmonics, and are shown in the bottom panel.}
    \label{fig:spec}
\end{figure*}

There are two exoplanet-hosting systems, HAT-P-2 \citep{2007ApJ...670..826B} and WASP-33 \citep{2010MNRAS.407..507C, 2011A&A...526L..10H}, for which it has been shown that the planets influence the pulsations of their host stars \citep{2017ApJ...836L..17D, 2022A&A...660L...2K}. In order to investigate this possibility, we computed the Fourier spectrum of the observed LC (Fig. \ref{fig:inputlc}, top panel) after subtracting the transits and the out-of-transit variations (Fig. \ref{fig:phase}). The frequency spectrum is shown in the top panel of Fig. \ref{fig:spec}. While a detailed study of the stellar oscillations is beyond to scope of this work, we suggest that HD 31221 is a $\delta$ Scuti/$\gamma$ Doradus hybrid pulsator according to the classification scheme of \cite{2010ApJ...713L.192G}.  For the frequency analysis, we used \texttt{Period04} \citep{2005CoAst.146...53L} and calculated the S/N of each frequency following the method used in \cite{1993A&A...271..482B}. We extracted 124 frequencies, adopting  $S/N$ > 4  as a criterion to distinguish between peaks due to pulsation and noise. The frequencies are  listed in Table \ref{tab:freqs}. The F2/F1 ratio is $0.61561$, which indicates that F1 is the fundamental mode and F2 is the second overtone \citep{1970ApJ...161..669F}, while the other frequencies are likely low amplitude non-radial modes. There are other $\delta$ Scuti stars with $>50$ identified frequencies in their light curves, including KIC 11754974 \citep{2013MNRAS.432.2284M} and KIC 10661783 \citep{2011MNRAS.414.2413S}.

The contaminating star, Gaia DR3 3412431471783190016, has $T_{\rm eff} = 7959.5 \pm 22.5$ K and $\log g = 4.144 \pm 0.012$ according to Gaia DR3. These parameters suggest that it could show $\delta$ Scuti type oscillations \citep{2011A&A...534A.125U}. Based on the brightness difference between HD 31221 and the contaminating star ($\Delta G = 5.08$ mag), the F1 and F2 peaks would correspond to a $\approx 0.68$ and $\approx 0.36$ Tmag pulsational amplitude, respectively. These are uncharacteristic even for so-called high-amplitude delta Scuti (HADS)\footnote{The pulsational amplitudes are usually computed in V-band, which is bluer than the TESS passband, and in which the pulsational amplitudes are even higher.} stars \citep{2009MNRAS.394..995D}, which are known to present only one high-amplitude pulsational frequency. However, in our analysis, even F3 has an amplitude of $\approx 0.27$ Tmag.  Although the light curves of Fig. \ref{fig:conta} are too noisy to extract high-quality Fourier spectra, they too suggest that HD 31221 is indeed a $\delta$ Scuti type star. Based on the available photometry, we are not able to eliminate the possibility that Gaia DR3 3412431471783190016 is also a star showing either $\delta$ Scuti or $\gamma$ Doradus pulsations, and thus it is possible that some of low-amplitude components of the Fourier spectra of Fig. \ref{fig:spec} are originating from it.

There are several instances where the orbital harmonics coincide with peaks in the spectrum. Three instances with the closest near-resonances ($3$rd, $85$th, and $221$st orbital harmonics, Table \ref{tab:freqs}) are shown on the bottom panels of Fig. \ref{fig:spec}. We note that these frequencies are present in the spectrum computed for the light curve shown in the upper panel of Fig. \ref{fig:conta}. To probe whether these resonance-like features are, in fact, a coincidence that may be attributed to  the large number of peaks present in the Fourier spectrum (Fig. \ref{fig:spec}), we constructed a simplified test. We created $10^6$ independent, randomized, uniform distributions of $124$ line segments each (corresponding to the number of frequencies extracted, Table \ref{tab:freqs}) between $0$ and $50$ d$^{-1}$ (where these frequencies can be found). As the width of the peaks in the original spectrum varies greatly, we set the width of every line segment at $0.006$ d$^{-1}$, which is equal to the distance between F12 and the $85$th orbital harmonic (Table \ref{tab:freqs}). We the calculated the total number of orbital harmonics that intersect these line segments in every distribution, allowing us to estimate the probability that a random uniform distribution can appear as the resonance-like features seen in Fig. \ref{fig:spec}. We found that in $2.7\%$ of the $10^6$ randomly generated distributions, there are three, four, or five orbital harmonics intersecting these synthetic peaks. We therefore suggest that there is tentative evidence that HD 31221 b influences the pulsation of its host, perhaps producing the so-called tidally perturbed oscillations observed in close-in binaries \citep{2020MNRAS.497L..19S, 2021PASJ...73..809L, 2021A&A...645A.119S} and the planet-hosting WASP-33 \cite{2022A&A...660L...2K}. Given that we have assumed a circular orbit for HD 31221 b (supported by the best-fit model covering the occultation well; see Fig. \ref{fig:phase}), we suggest that the tidal forces arising from the giant planet or brown dwarf on a misaligned orbit are responsible for influencing the stellar oscillations.

\section{Summary and conclusion}

We present the discovery of a substellar companion to HD 31221 b. Given that there is only one star near the target (Fig. \ref{fig:tpf}), which is between four and six magnitudes dimmer than the target, it is extremely unlikely that the transits are the result of a background eclipsing binary. HD 31221 b has a radius of $1.32 \pm 0.14$~R$_J$ with an orbital period of $4.66631 \pm 0.00011$ days. By modeling the ellipsoidal variations, we find a mass of mass of $11.5 \pm 10.3$~M$_J$. By modeling the Doppler beaming, we can place a $1-\sigma$ upper constraint on its mass of $26.3$~M$_J$, suggesting that it is either a hot Jupiter or a brown dwarf. Through modeling the gravity darkening, HD 31221 b is found to have a misaligned orbit, with an obliquity of $-121.6 \pm 14.4^\circ$. Therefore HD 31221 b  joins WASP-33 \citep{2011A&A...526L..10H} and KOI-976 \citep{2019AJ....158...88A} as a type of substellar object orbiting a star presenting $\delta$ Scuti oscillations and gravity darkening, with a short orbital period.

In this system, the out-of-transit variations (see Fig. \ref{fig:phase}) appear to be dominated by the reflection effect, appearing in our model as a high geometric albedo of $1.58 \pm 0.50$. A non-zero geometric albedo was also found in the case of KELT-1 b \citep{2020AJ....160..211B}, namely, another brown with a short orbital period ($1.21749394 \pm 2.5 \cdot 10^{-7}$ days). Because of the degeneracies between this and the beaming effect, we are unable to precisely determine either. Given that radial velocity measurements are not a plausible way to estimate the mass of HD 31221 b, simultaneous analyses of multi-color photometric observations (with data from CHEOPS or PLATO \citep{2014ExA....38..249R}) can be used to have better mass and albedo estimates. By imposing a $\mathcal{N} (0.5,0.1)$ Gaussian prior on the geometric albedo, we get slightly higher mass estimates ($13.0 \pm 11.1$ and $15.6 \pm 14.2$~M$_J$ from the ellipsoidal variations and the Doppler beaming, respectively). Because of the large uncertainties, all mass estimates agree well with each other. We measured a nightside flux that is consistent with $0$, and an offset between the brightest point of the companions surface and the substellar point of $52.6 \pm 41.2^\circ$, suggestive of clouds in the atmosphere of HD 31221 b. We found the secondary occultation depth to be $99 \pm 9$ ppm.


By analyzing the Fourier spectrum of the pulsations of HD 31221, we found evidence that HD 31221 is tidally influencing the pulsations of its host, similarly to the case of WASP-33 \citep{2022A&A...660L...2K}. Figure \ref{fig:spec} displays the $3$rd, $85$th, and $221$st orbital harmonics, showing that there are nearly exact resonances with the pulsational frequencies.

\begin{acknowledgements} We thank L. Moln\'ar for helpful discussions about Gaia DR3. This work has made use of the TIC, through the TESS Science Office’s target selection working group (architects K. Stassun, J. Pepper, N. De Lee, M. Paegert, R. Oelkers). The Filtergraph data portal system is trademarked by Vanderbilt University. This work was supported by the PRODEX Experiment Agreement No. 4000137122 between the ELTE E\"otv\"os Lor\'and University and the European Space Agency (ESA-D/SCI-LE-2021-0025). Support of the Lend\"ulet LP2018-7/2022 grant of the Hungarian Academy of Science, and the KKP-137523 `SeismoLab' \'Elvonal  grant  as well as the grant K-138962 of the Hungarian Research, Development and Innovation Office (NKFIH) are acknowledged. Project no. C1746651 has been implemented with the support provided by the Ministry of Culture and Innovation of Hungary from the National Research, Development and Innovation Fund, financed under the NVKDP-2021 funding scheme.
     
\end{acknowledgements}

%
%
\bibliographystyle{aa}
\bibliography{refs}

\begin{appendix}

\section{Gravity-darkened transit}

\begin{figure}[!h]
    \centering
    \includegraphics[width = \columnwidth]{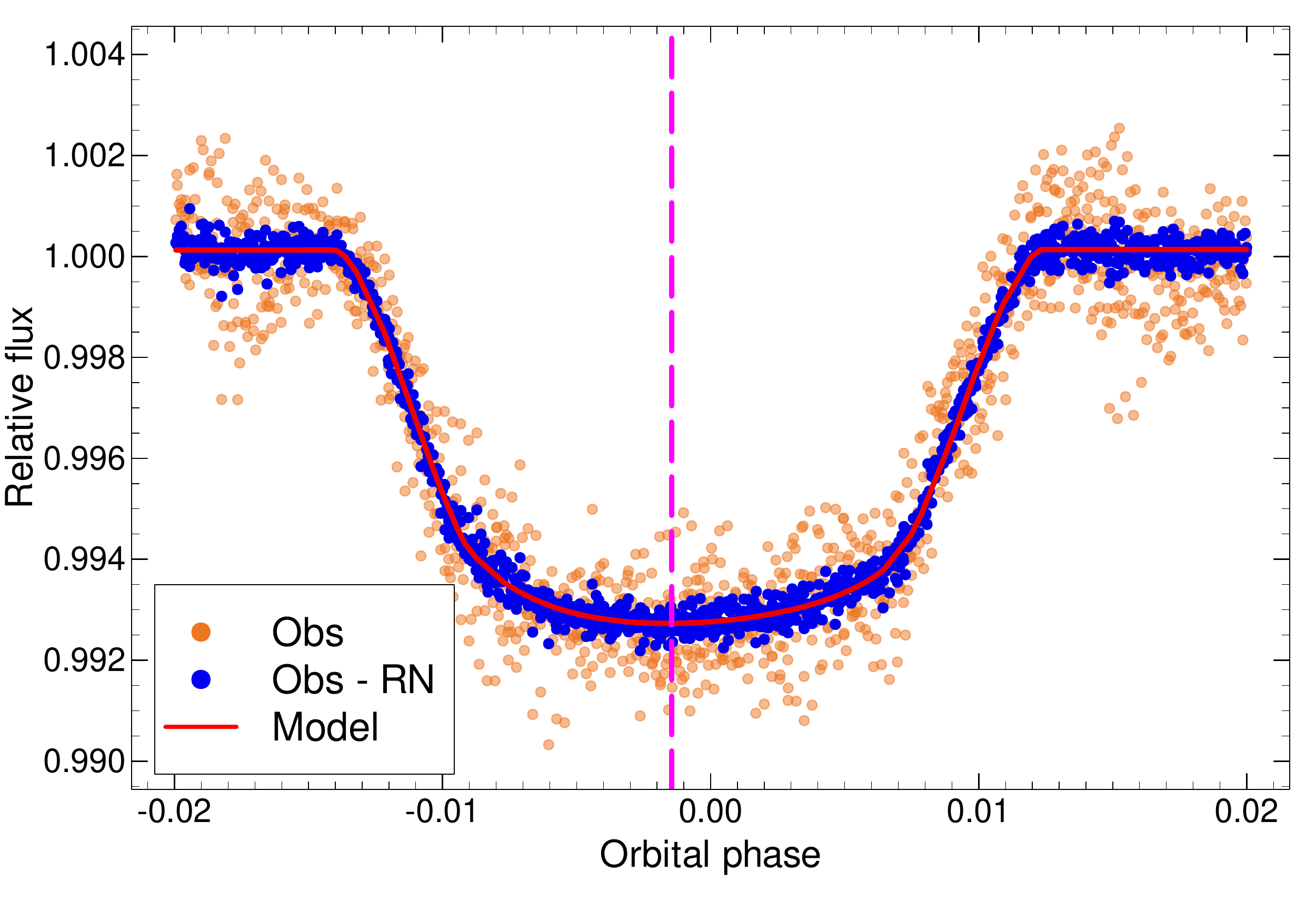}
    \caption{Phase folded, pre-cleaned light curve of HD 31221 b (orange dots), centered around the transit. Removing the correlated noise from the input LC yield the blue dots, while the solid red line represents the best-fit solution. The dashed magenta line is drawn at the minimum of the transit, which is shifted from $0.0$ phase because of gravity darkening.}
    \label{fig:gravdark}
\end{figure}


\section{HD 31221 as the source of the transits and the phase curve}

\begin{figure}[!h]
    \centering
    \includegraphics[width = 8 cm]{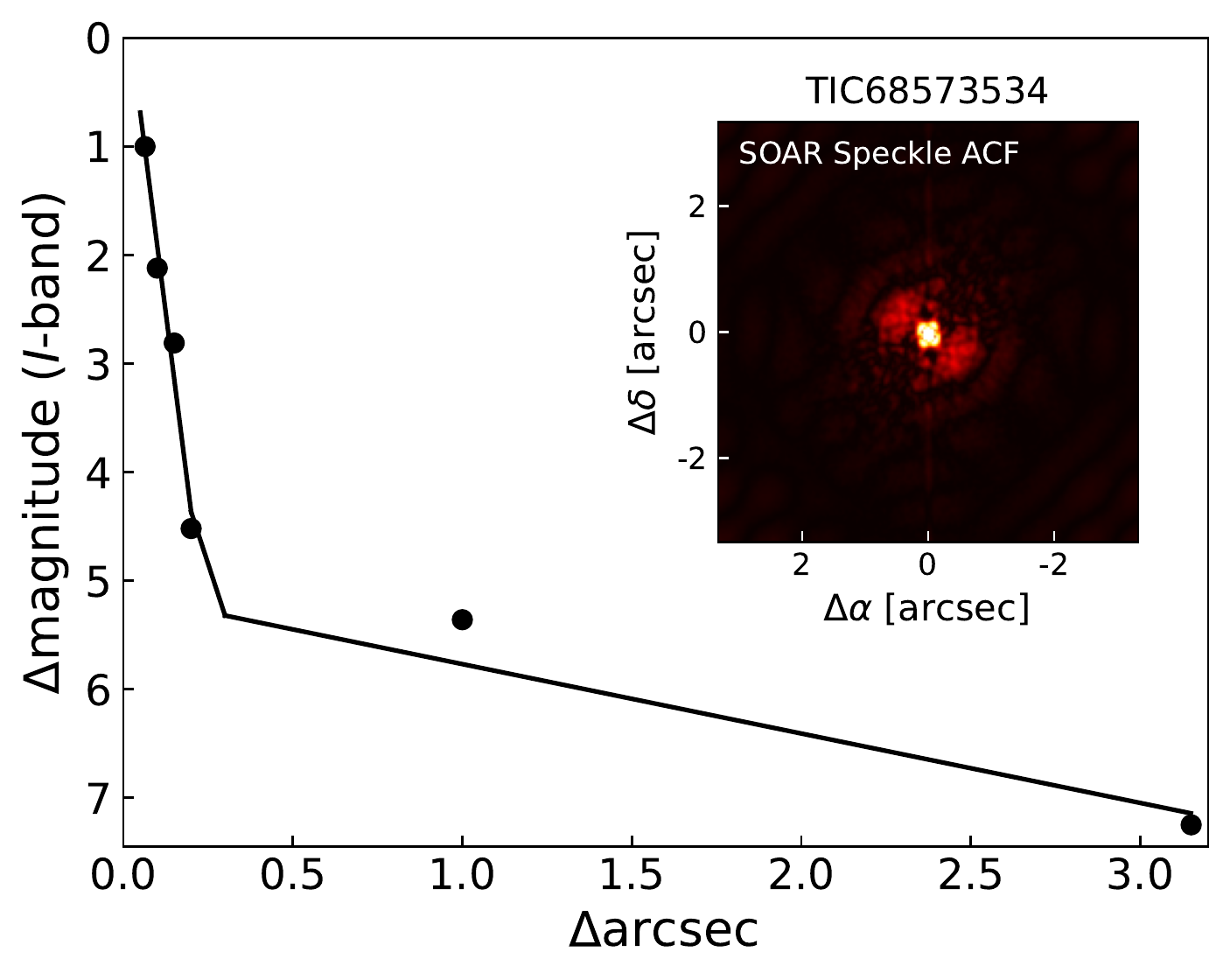}
    \caption{Sensitivity curve and auto-correlation function from the speckle imaging with SOAR in $I$-band, which is similar to the TESS passband.}
    \label{fig:soar}
\end{figure}

\begin{figure}[!h]
    \centering
    \includegraphics[width = .85\columnwidth]{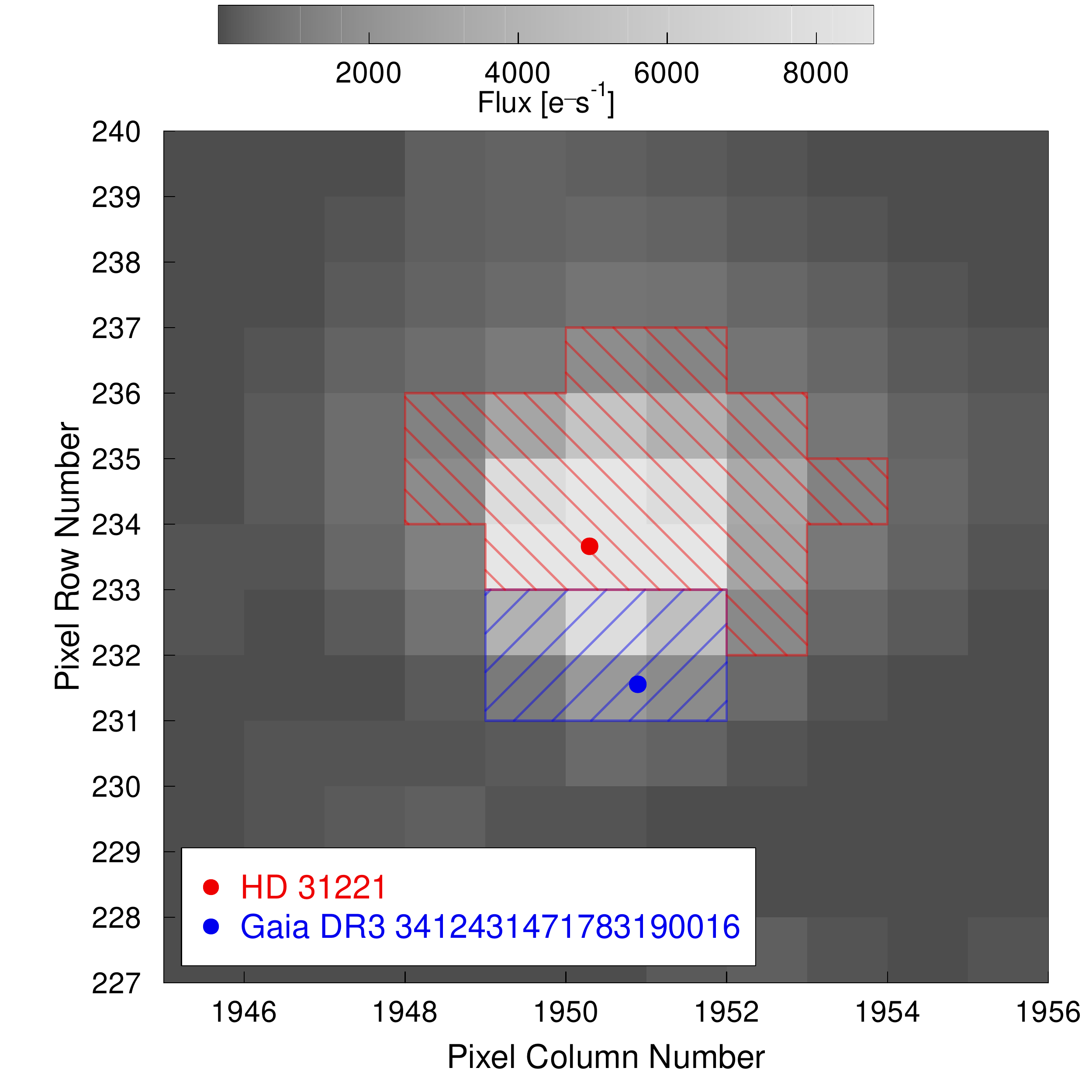}
    \caption{Custom apertures used to separate the PSF of the HD 31221 (red) and Gaia DR3 3412431471783190016 (blue) in Sector 43. Red and blue dots mark the approximate pixel coordinates of the two stars. The corresponding light curves are shown in Fig. \ref{fig:conta}.}
    \label{fig:aps}
\end{figure}

\begin{figure}[!h]
    \centering
    \includegraphics[width = .95\columnwidth]{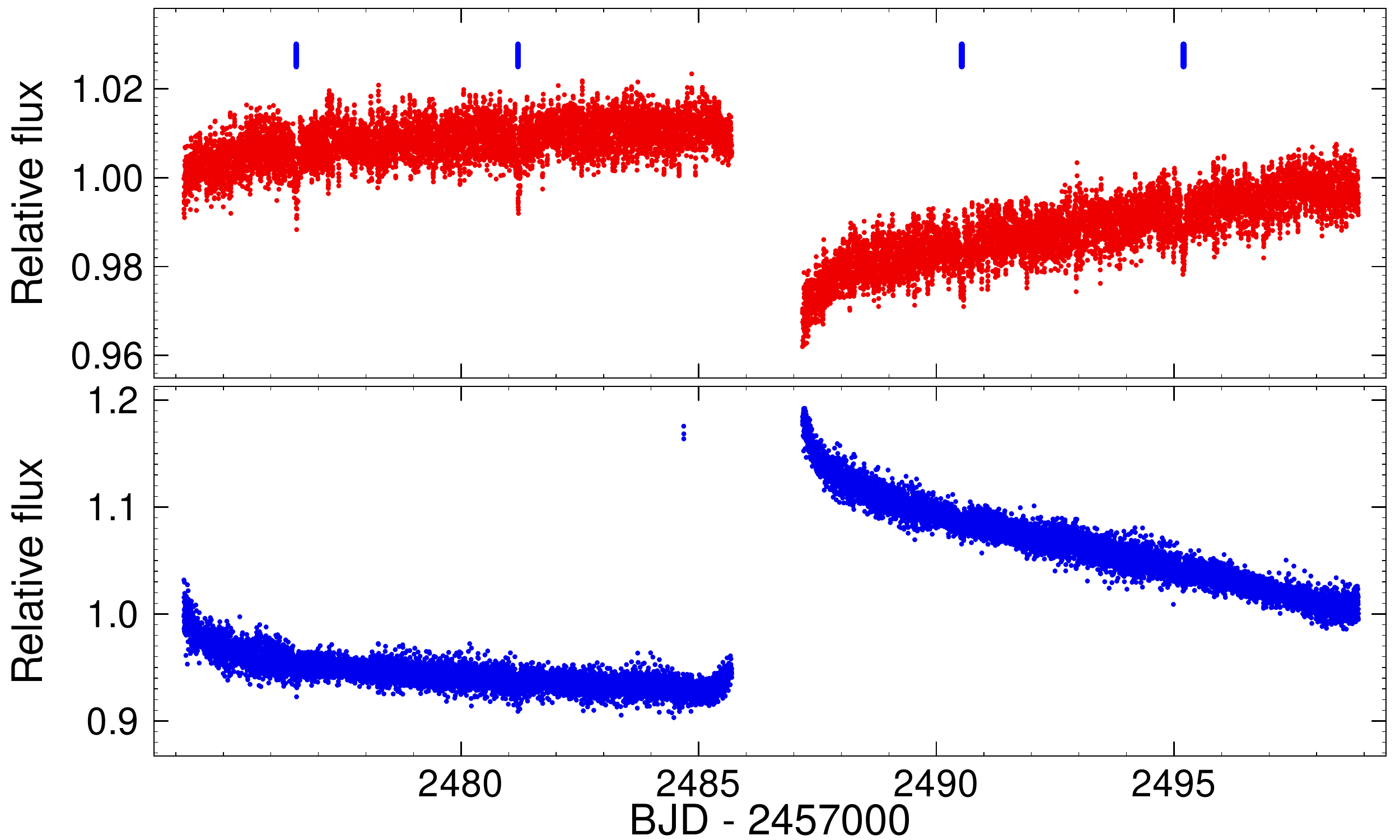}
    \caption{Light curves from Sector 43, extracted by the custom apertures (Fig. \ref{fig:aps}) for the photometry of HD 31221 (top panel) and Gaia DR3 3412431471783190016 (bottom panel). Blue ticks mark the transits seen in Fig. \ref{fig:inputlc}.}
    \label{fig:conta}
\end{figure}

\begin{figure}
    \centering
    \includegraphics[width = \columnwidth]{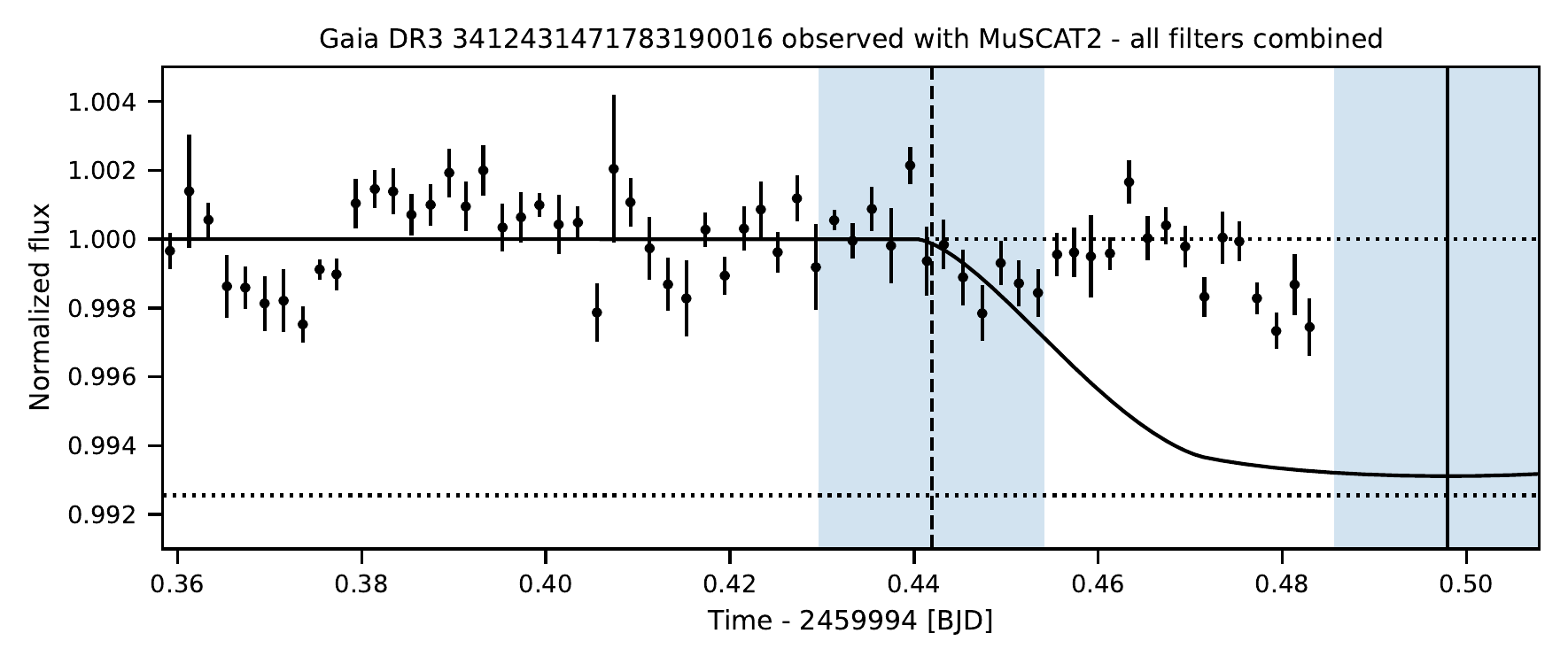}
        \caption{Light curve of Gaia DR3 3412431471783190016 covering the expected transit ingress observed with MuSCAT2. The dots with uncertainties show the MuSCAT2 observations, the solid line visualises the transit, the solid vertical line shows the expected transit center, and the slashed vertical line shows the expected beginning of the transit. The observations clearly show that the transit signal does not arise from Gaia DR3 3412431471783190016.}
    \label{fig:muscat-gaia}
\end{figure}

We performed speckle imaging (Fig. \ref{fig:soar}), defined custom apertures (Fig. \ref{fig:aps}), extracted the respective light curves (Fig. \ref{fig:conta}), and conducted ground-based observations (Fig. \ref{fig:muscat-gaia}) in order to confirm that HD 31221 is the source of the observed signal (Fig. \ref{fig:phase}).

 \clearpage
\section{Posteriors of the MCMC sampling}
\begin{figure*}
    \centering
    \includegraphics[width = \textwidth]{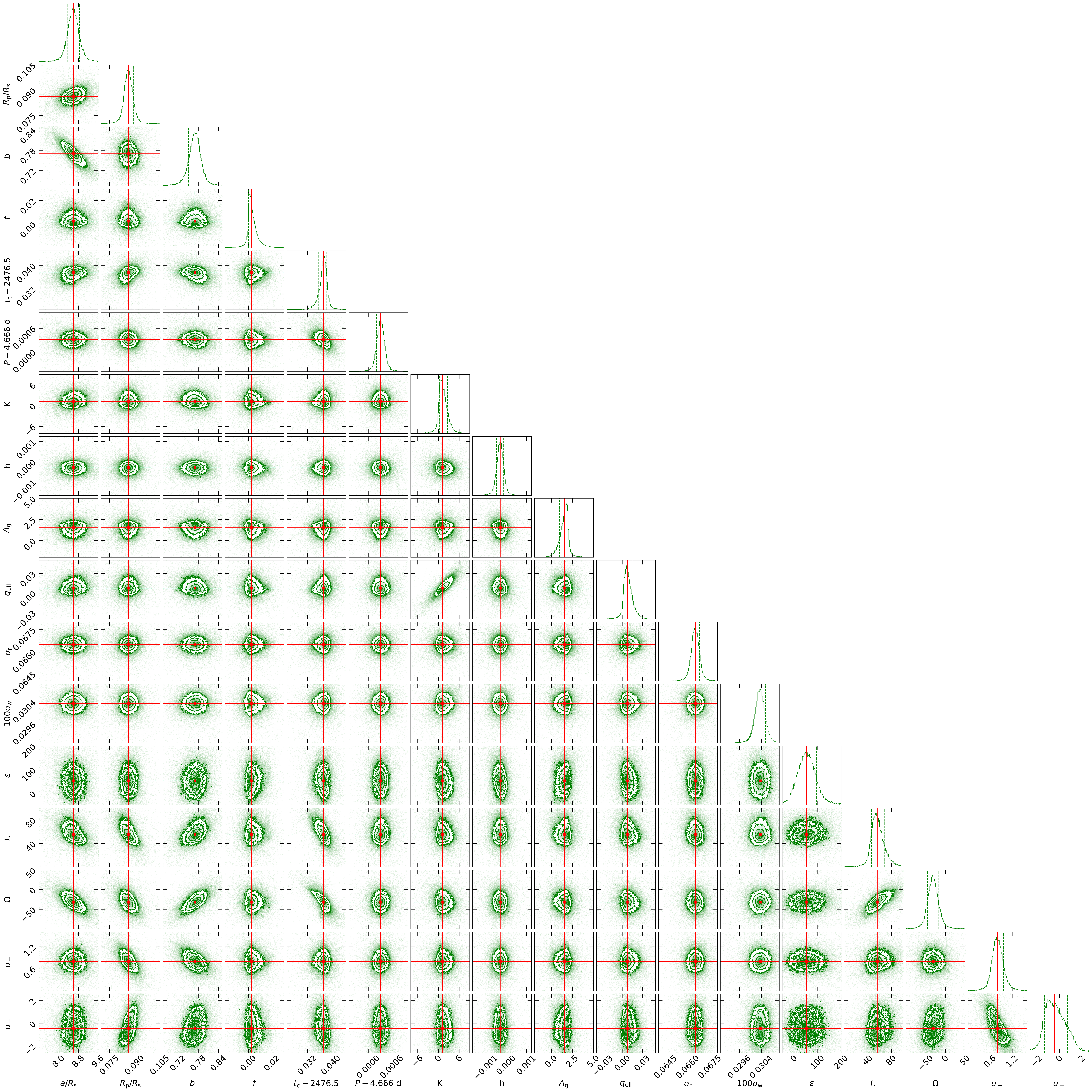}
    \caption{Corner plot created with the routines of \cite{corner},  showing the analysis of the free $A_g$ case.}
    \label{fig:corner}
\end{figure*}
 \onecolumn

\section{List of extracted frequencies}
\footnotesize
\begin{longtable}{l c c c c c c c}

\caption{\label{tab:freqs} Extracted frequencies and the nearest orbital $n$th orbital harmonics from the Fourier spectrum observed light curve of HD 31221, with the signal directly attributed to HD 31221 b subtracted.}\\
\hline \hline
ID & Frequency [d$^{-1}$] & Amplitude [ppm] & Phase [rad] & S/N & $n$ & $n \cdot P^{-1}$ [d$^{-1}$] & Frequency $-n \cdot P^{-1}$ [d$^{-1}$] \\
\hline
\endfirsthead
\caption{continued.}\\
\hline\hline
Order of frequencies & Frequency [d$^{-1}$] & Amplitude [ppm] & Phase [rad] & SNR & $n$ & $n \cdot P^{-1}$ [d$^{-1}$] & Frequency $-n \cdot P^{-1}$ [d$^{-1}$] \\ 
\hline
\endhead
\hline
\endfoot
F1 & $31.70744$ & $2712.6$ & $0.9023$ & $179.3$ & & & \\
F2 & $19.51938$ & $1490.8$ & $0.2270$ & $124.9$ & & & \\
F3 & $21.23275$ & $1140.0$ & $0.9357$ & $66.1$ & & & \\
F4 & $23.06200$ & $1068.8$ & $0.3939$ & $68.4$ & & & \\
F5 & $37.55911$ & $1049.9$ & $0.8990$ & $70.2$ & & & \\
F6 & $22.05100$ & $964.7$ & $0.7681$ & $55.6$ & & & \\
F7 & $17.59203$ & $903.6$ & $0.6996$ & $68.6$ & & & \\
F8 & $29.13693$ & $770.6$ & $0.1748$ & $58.1$ & & & \\
F9 & $0.64234$ & $738.1$ & $0.9602$ & $28.2$ & $3$ & $0.64291$ & $-0.0057$\\
F10 & $22.37120$ & $722.3$ & $0.6171$ & $44.6$ & & & \\
F11 & $24.01671$ & $590.2$ & $0.4124$ & $40.5$ & & & \\
F12 & $18.22191$ & $607.5$ & $0.5354$ & $47.1$ & $85$ & $18.21568$ & $0.00611$\\
F13 & $43.56254$ & $538.1$ & $0.7799$ & $38.6$ & & & \\
F14 & $47.27522$ & $497.5$ & $0.0971$ & $27.4$ & & & \\
F15 & $18.38763$ & $504.2$ & $0.8846$ & $40.0$ & & & \\
F16 & $30.81062$ & $489.4$ & $0.3908$ & $30.3$ & & & \\
F17 & $51.14577$ & $489.4$ & $0.6115$ & $25.4$ & & & \\
F18 & $35.24569$ & $486.8$ & $0.0455$ & $37.0$ & & & \\
F19 & $22.94561$ & $421.9$ & $0.5800$ & $27.4$ & & & \\
F20 & $29.23169$ & $409.0$ & $0.1423$ & $29.9$ & & & \\
F21 & $50.09471$ & $415.3$ & $0.8169$ & $25.0$ & & & \\
F22 & $41.32154$ & $375.2$ & $0.4344$ & $19.4$ & & & \\
F23 & $24.08516$ & $359.6$ & $0.6613$ & $24.5$ & & & \\
F24 & $43.97839$ & $344.3$ & $0.4799$ & $21.9$ & & & \\
F25 & $18.19144$ & $318.9$ & $0.5552$ & $24.4$ & & & \\
F26 & $21.64876$ & $289.6$ & $0.7562$ & $16.0$ & & & \\
F27 & $0.07021$ & $810.1$ & $0.3018$ & $31.7$ & & & \\
F28 & $51.68677$ & $279.2$ & $0.2626$ & $16.5$ & & & \\
F29 & $47.36391$ & $257.4$ & $0.0501$ & $14.2$ & $221$ & $47.36076$ & $0.00315$ \\
F30 & $40.79371$ & $276.0$ & $0.5573$ & $15.0$ & & & \\
F31 & $41.02223$ & $248.1$ & $0.3904$ & $12.9$ & & & \\
F32 & $3.06930$ & $247.1$ & $0.8694$ & $12.6$ & & & \\
F33 & $38.40938$ & $249.2$ & $0.9867$ & $16.6$ & & & \\
F34 & $22.03629$ & $292.9$ & $0.3693$ & $16.9$ & & & \\
F35 & $56.96805$ & $216.8$ & $0.0241$ & $12.7$ & & & \\
F36 & $24.82204$ & $210.7$ & $0.1266$ & $13.8$ & & & \\
F37 & $45.44228$ & $211.5$ & $0.8337$ & $15.3$ & & & \\
F38 & $37.09706$ & $244.8$ & $0.7629$ & $15.6$ & & & \\
F39 & $0.45097$ & $201.5$ & $0.8532$ & $7.8$ & & & \\
F40 & $34.11424$ & $198.0$ & $0.5890$ & $14.7$ & & & \\
F41 & $50.59839$ & $207.0$ & $0.6475$ & $10.8$ & & & \\
F42 & $35.15254$ & $198.1$ & $0.1131$ & $15.0$ & & & \\
F43 & $0.03600$ & $274.4$ & $0.8555$ & $10.7$ & & & \\
F44 & $39.04703$ & $191.0$ & $0.5965$ & $13.7$ & & & \\
F45 & $51.70983$ & $237.3$ & $0.7599$ & $14.1$ & & & \\
F46 & $18.37027$ & $199.5$ & $0.9437$ & $15.6$ & & & \\
F47 & $28.96813$ & $192.1$ & $0.9105$ & $14.8$ & & & \\
F48 & $53.46973$ & $162.3$ & $0.9821$ & $10.2$ & & & \\
F49 & $18.47889$ & $178.3$ & $0.9109$ & $14.2$ & & & \\
F50 & $27.84964$ & $166.4$ & $0.2797$ & $13.9$ & & & \\
F51 & $0.14053$ & $199.6$ & $0.4713$ & $7.8$ & & & \\
F52 & $8.70011$ & $164.5$ & $0.6195$ & $10.7$ & & & \\
F53 & $33.70630$ & $173.3$ & $0.6495$ & $12.5$ & & & \\
F54 & $18.99892$ & $179.1$ & $0.6324$ & $13.5$ & & & \\
F55 & $53.81707$ & $171.1$ & $0.8036$ & $10.4$ & & & \\
F56 & $22.39475$ & $167.2$ & $0.1219$ & $10.3$ & & & \\
F57 & $32.94004$ & $160.2$ & $0.2262$ & $11.5$ & & & \\
F58 & $37.17270$ & $161.7$ & $0.6523$ & $10.2$ & & & \\
F59 & $18.97378$ & $164.1$ & $0.6779$ & $12.4$ & & & \\
F60 & $26.53367$ & $155.2$ & $0.6862$ & $11.7$ & & & \\
F61 & $44.24069$ & $150.5$ & $0.8335$ & $9.6$ & & & \\
F62 & $44.06767$ & $142.2$ & $0.5932$ & $9.2$ & & & \\
F63 & $44.79974$ & $141.4$ & $0.3872$ & $9.6$ & & & \\
F64 & $0.07365$ & $653.7$ & $0.2607$ & $25.6$ & & & \\
F65 & $28.62958$ & $132.7$ & $0.1874$ & $10.4$ & & & \\
F66 & $3.22543$ & $136.7$ & $0.9299$ & $7.1$ & & & \\
F67 & $34.33265$ & $142.4$ & $0.9892$ & $10.5$ & & & \\
F68 & $35.70868$ & $130.3$ & $0.9897$ & $9.5$ & & & \\
F69 & $43.26799$ & $118.5$ & $0.1595$ & $7.7$ & & & \\
F70 & $24.56078$ & $119.1$ & $0.5059$ & $7.5$ & & & \\
F71 & $7.20946$ & $128.8$ & $0.3198$ & $8.4$ & & & \\
F72 & $25.70421$ & $129.8$ & $0.5053$ & $8.4$ & & & \\
F73 & $1.13105$ & $129.9$ & $0.9883$ & $4.8$ & & & \\
F74 & $21.75384$ & $137.6$ & $0.8991$ & $7.9$ & & & \\
F75 & $47.20039$ & $124.4$ & $0.8566$ & $7.2$ & & & \\
F76 & $26.85858$ & $125.1$ & $0.4142$ & $9.4$ & & & \\
F77 & $47.31264$ & $120.2$ & $0.9492$ & $6.6$ & & & \\
F78 & $30.96470$ & $118.1$ & $0.4663$ & $7.4$ & & & \\
F79 & $53.75625$ & $111.3$ & $0.7334$ & $6.9$ & & & \\
F80 & $29.85313$ & $113.9$ & $0.3917$ & $7.6$ & & & \\
F81 & $0.28845$ & $117.9$ & $0.7971$ & $4.5$ & & & \\
F82 & $29.42938$ & $113.4$ & $0.0691$ & $8.2$ & & & \\
F83 & $13.76900$ & $110.9$ & $0.1340$ & $7.6$ & & & \\
F84 & $27.62043$ & $108.5$ & $0.8008$ & $8.5$ & & & \\
F85 & $25.28139$ & $107.1$ & $0.4505$ & $6.9$ & & & \\
F86 & $20.05024$ & $104.4$ & $0.6098$ & $7.4$ & & & \\
F87 & $1.34457$ & $106.5$ & $0.0397$ & $4.0$ & & & \\
F88 & $7.34244$ & $99.7$ & $0.1219$ & $6.7$ & & & \\
F89 & $20.59324$ & $98.9$ & $0.0672$ & $5.7$ & & & \\
F90 & $32.54621$ & $100.5$ & $0.3924$ & $6.9$ & & & \\
F91 & $40.99852$ & $99.7$ & $0.0660$ & $5.2$ & & & \\
F92 & $2.84827$ & $96.8$ & $0.5668$ & $4.7$ & & & \\
F93 & $50.61361$ & $103.7$ & $0.3200$ & $5.4$ & & & \\
F94 & $3.32463$ & $95.8$ & $0.1075$ & $5.4$ & & & \\
F95 & $38.81757$ & $92.6$ & $0.4319$ & $6.1$ & & & \\
F96 & $1.74337$ & $95.7$ & $0.3033$ & $3.6$ & & & \\
F97 & $40.77376$ & $99.3$ & $0.5802$ & $5.4$ & & & \\
F98 & $37.04735$ & $96.6$ & $0.5899$ & $6.3$ & & & \\
F99 & $42.57282$ & $93.2$ & $0.6170$ & $6.0$ & & & \\
F100 & $43.23417$ & $101.3$ & $0.1742$ & $6.7$ & & & \\
F101 & $43.54843$ & $107.8$ & $0.7735$ & $7.6$ & & & \\
F102 & $32.88931$ & $95.1$ & $0.2398$ & $6.6$ & & & \\
F103 & $24.06407$ & $91.6$ & $0.5862$ & $6.3$ & & & \\
F104 & $27.36682$ & $91.6$ & $0.5218$ & $7.5$ & & & \\
F105 & $51.65884$ & $97.9$ & $0.9135$ & $5.6$ & & & \\
F106 & $53.44199$ & $96.9$ & $0.7775$ & $6.1$ & & & \\
F107 & $26.71840$ & $90.4$ & $0.7146$ & $7.0$ & & & \\
F108 & $2.52605$ & $91.7$ & $0.5020$ & $4.0$ & & & \\
F109 & $25.34499$ & $91.1$ & $0.8186$ & $5.9$ & & & \\
F110 & $45.06505$ & $87.2$ & $0.9064$ & $5.9$ & & & \\
F111 & $26.05606$ & $90.0$ & $0.4267$ & $5.9$ & & & \\
F112 & $21.69018$ & $89.2$ & $0.9519$ & $5.0$ & & & \\
F113 & $34.85446$ & $87.4$ & $0.1269$ & $6.6$ & & & \\
F114 & $38.69624$ & $87.0$ & $0.6495$ & $5.6$ & & & \\
F115 & $37.62416$ & $87.3$ & $0.6848$ & $5.6$ & & & \\
F116 & $40.35308$ & $83.6$ & $0.5058$ & $5.1$ & & & \\
F117 & $0.62753$ & $90.8$ & $0.7177$ & $3.4$ & & & \\
F118 & $48.73080$ & $82.0$ & $0.1627$ & $5.0$ & & & \\
F119 & $34.26273$ & $86.1$ & $0.8970$ & $6.3$ & & & \\
F120 & $26.11275$ & $78.5$ & $0.1955$ & $5.2$ & & & \\
F121 & $47.09981$ & $78.8$ & $0.3535$ & $4.7$ & & & \\
F122 & $25.24455$ & $77.9$ & $0.6486$ & $5.0$ & & & \\
F123 & $24.60607$ & $79.2$ & $0.2934$ & $5.0$ & & & \\
F124 & $8.93464$ & $78.0$ & $0.0155$ & $5.0$ & & & \\
\hline
\end{longtable}

\end{appendix}
\end{document}